# Fast Nonparametric Conditional Density Estimation

**Michael P. Holmes**
College of Computing
Georgia Institute of Technology
Atlanta, GA 30332 USA
mph@cc.gatech.edu

**Alexander G. Gray**
College of Computing
Georgia Institute of Technology
Atlanta, GA 30332 USA
agray@cc.gatech.edu

**Charles Lee Isbell, Jr.**
College of Computing
Georgia Institute of Technology
Atlanta, GA 30332 USA
isbell@cc.gatech.edu

## Abstract

Conditional density estimation generalizes regression by modeling a full density $f(y|x)$ rather than only the expected value $E(y|x)$. This is important for many tasks, including handling multi-modality and generating prediction intervals. Though fundamental and widely applicable, nonparametric conditional density estimators have received relatively little attention from statisticians and little or none from the machine learning community. None of that work has been applied to greater than bivariate data, presumably due to the computational difficulty of data-driven bandwidth selection. We describe the double kernel conditional density estimator and derive fast dual-tree-based algorithms for bandwidth selection using a maximum likelihood criterion. These techniques give speedups of up to 3.8 million in our experiments, and enable the first applications to previously intractable large multivariate datasets, including a redshift prediction problem from the Sloan Digital Sky Survey.

## 1 Introduction

Conditional density estimation is the estimation of the probability density $f(y|x)$ of a random variable $y$ given a random vector $x$. For example, in Figure 1 each contour line perpendicular to the $x$ axis represents a conditional density. This can be viewed as a generalization of regression: in regression we estimate the expectation $E[y|x]$, while in conditional density estimation we model the full distribution. Figure 1 illustrates a conditional bimodality for which $E[y|x]$ is insufficiently descriptive. Estimating conditional densities is much harder than regression, but having the full distribution is powerful because it allows one to extract almost any quantity of interest, including the expectation, modes, prediction intervals, outlier boundaries, samples, expectations of non-linear functions of $y$, etc. It also facilitates data visualization and exploration. Conditional density estimates are of fundamental and widespread utility, and are applicable to such problems as nonparametric continuous Markov models, nonparametric estimation of conditional distributions within Bayes nets, time series prediction, and static regression with prediction intervals. The estimation problem is challenging because the data generally do not include the exact $x$ for which $f(y|x)$ is desired. Nonparametric kernel techniques address this issue by interpolating between the points we have seen without making distributional assumptions.

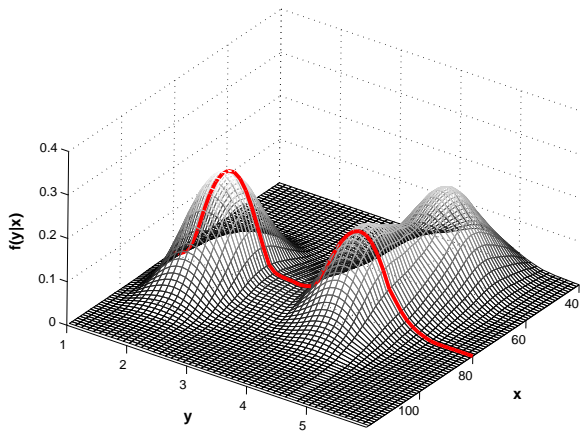

Figure 1: Dataset for which $f(y|x)$ can be either bimodal or unimodal, depending on $x$. The bold curve represents $f(y|x = 80)$.

In nonparametric conditional density estimation, we make only minimal assumptions about the smoothness of $f(y|x)$ without assuming any parametric form. Freedom from parametric assumptions is very often desirable when dealing with complex data, as we rarely have knowledge of true distributional forms. While a small amount of work on nonparametric kernel conditional density estimation has been done by statisticians and econometrics researchers (Gooijer & Zerom, 2003; Fan & Yim, 2004; Hansen, 2004; Bashtannyk & Hyndman, 2001; Hyndman et al., 1996; Rosenblatt, 1969), it appears to have received little or no attention from the machine learning community. Note that what we mean by nonparametric conditional density estimation is different from other machine learning techniques with similar names, such as conditional probability estimation (which refers to outputting class probabilities in the classification setting, also referred to as class-conditional probabilities) and various discrete and/or parametric conditional density models such as those commonly used in Bayes nets. The only machine learning work we have found that seems to look at the same problem is (Schapire et al., 2002), but it employs a discretization scheme rather than handling continuous values directly.

In the present work, we use the standard kernel conditional density estimator that first received serious attention in (Fan et al., 1996) and (Hyndman et al., 1996), though it was originally proposed in (Rosenblatt, 1969). Although this estimator is consistent given mild conditions on its bandwidths, practical use has been hampered by the lack of an efficient data-driven bandwidth selection procedure, upon which any kernel estimator depends critically. We propose a new method for efficiently selecting bandwidths to maximize cross-validated likelihood. The speedup of this method is obtained via a dual-tree-based approximation (Gray & Moore, 2000) of the likelihood function. Speeding up likelihood evaluations is relevant for general nonparametric inference, but the present work focuses on its application to bandwidth selection. We present two versions of likelihood approximation, one analogous to previous dual-tree algorithms with deterministic error control, which gives speedups as high as 667 on our datasets, and the other with a new sampling-based probabilistic error control mechanism, giving much larger speedups as high as 3.8 million.

With this fast inference procedure we are able to address datasets of greater dimensionality and an order of magnitude larger than in previous work, which was confined to bivariate datasets of size no greater than 1000 (Fan & Yim, 2004). We present results that validate the accuracy and speedup of our likelihood approximation on datasets possessing a variety of sizes and dimensionalities. We also present results on the quality of the resulting density estimates and their predictions on various synthetic datasets (which allow us to compare to known distributions) and on a Sloan Digital Sky Survey (SDSS) redshift prediction problem of current scientific interest. Most of these datasets were previously unaddressable by naively-computed data-driven techniques. Our kernel estimators perform well compared to a standard reference rule bandwidth procedure on all datasets, and, though not designed for regression, are competitive in terms of regression metrics with the de facto algorithm employed by astronomers on the SDSS dataset. We conclude that kernel conditional density estimation is a powerful technique that is made substantially more efficient by our fast inference procedure, with many opportunities for application in machine learning.

## 2 Kernel conditional density estimation

In unconditional kernel density estimation (KDE), we estimate a probability distribution $f(x)$ from a dataset $\{x_i\}$ by $\hat{f}(x) = \frac{1}{n}\sum_i K_h(||x - x_i||)$, where $K_h(t) = \frac{1}{h^d}K(\frac{t}{h})$, $K$ is a kernel function, i.e. a compact, symmetric probability distribution such as the Gaussian or Epanechnikov, $d$ is the dimension of $x$, $n$ is the number of data points, and $h$ is the bandwidth controlling the kernel widths (see Silverman, 1986). Kernels allow us to interpolate between the data we have seen in order to predict the density at points we haven't seen.

In kernel conditional density estimation (KCDE), this interpolation must happen in both the $x$ and $y$ directions, which leads to a double kernel estimator:

$$\hat{f}(y|x) = \frac{\sum_i K_{h_1}(y - y_i) K_{h_2}(||x - x_i||)}{\sum_i K_{h_2}(||x - x_i||)} \ . \quad (1)$$

This form is known as the Nadaraya-Watson (NW) conditional density estimator (Gooijer & Zerom, 2003). For a queried $x$, it constructs a density by weighting each $y_i$ proportionally to the proximity of the corresponding $x_i$. Figure 1 illustrates a conditional density estimate on a dataset with univariate $x$.

The NW estimator is consistent provided $h_1 \to 0$, $h_2 \to 0$, and $nh_1h_2 \to \infty$ as $n \to \infty$ (Hyndman et al., 1996). A few statisticians and econometrics researchers have made extensions to the NW estimator, most notably by the addition of local polynomial smoothing (Fan et al., 1996; Fan & Yim, 2004; Gooijer & Zerom, 2003). They have also proposed both reference rules and data-driven bandwidth selection procedures, but all applications appear to have been confined to the bivariate case, as has most of the theoretical analysis. One likely reason for this limitation is the difficulty of selecting good bandwidths in the presence of large datasets and higher dimensionality.

## 3 Bandwidth selection

As with all kernel estimators, the performance of the NW estimator depends critically on a suitable choice for the bandwidths $h_1$ and $h_2$. The aforementioned consistency conditions provide little guidance in the finite-sample setting. Bandwidth selection has always been a dilemma: on the one hand, asymptotic arguments and reference distributions lead to plug-in and reference rules whereby bandwidths can be efficiently calculated, but these perform poorly on finite samples and when reference distributions don't match reality; on the other hand, data-driven selection criteria give good bandwidths but are naively intractable on datasets of appreciable size. We propose a middle road that captures some of the advantage of each approach by generating efficient approximations to the naively expensive data-driven computations.

The only data-driven bandwidth score to previously appear in the KCDE literature is the integrated squared error in the following form:

$$ISE(h_1, h_2) = \int (f(y|x) - \hat{f}(y|x))^2 dy f(x) dx \ . \quad (2)$$

As shown in (Fan & Yim, 2004), minimizing ISE is equivalent to minimizing $\int (\hat{f}(y|x))^2 dy f(x) dx - 2 \int \hat{f}(y|x) f(y, x) dy dx$. A consistent, cross-validated estimate of the ISE is obtained by $\widehat{ISE} = \frac{1}{n} \sum_i \int (\hat{f}^{-i}(y|x_i))^2 dy - \frac{2}{n} \sum_i \hat{f}^{-i}(y_i|x_i)$, where $\hat{f}^{-i}$ denotes $\hat{f}$ evaluated with $(x_i, y_i)$ left out.

Though appealing, the first term of $\widehat{ISE}$ expands to a triply-nested summation, giving a base computational cost of $O(n^3)$. While this could still be used as the starting point for an efficient approximation, we choose to start with another criterion that has lower base complexity: likelihood cross-validation.

Likelihood cross-validation has long been known in standard kernel density estimation (see Silverman, 1986; Gray & Moore, 2003), but has yet to be used for KCDE. One likely reason for this is the non-robustness to outliers that can afflict the likelihood function, particularly in the presence of heavy-tailed distributions. Although well-known asymptotic results motivate the use of ISE instead of likelihood, we turn to the likelihood for this problem because of the significant computational benefit, balanced by our empirical observation that its performance is close to that of ISE.

By analogy with (Silverman, 1986), we define the cross-validated log likelihood for KCDE to be:

$$L(h_1, h_2) = \frac{1}{n} \sum_i \log(\hat{f}^{-i}(y_i|x_i)\hat{f}^{-i}(x_i)) \ , \quad (3)$$

where $\hat{f}(x)$ is the standard kernel density estimate over $x$ using the bandwidth $h_2$ from $\hat{f}(y|x)$. We want to choose the bandwidth pair $(h_1, h_2)$ that minimizes $-L$; by so doing, we will be minimizing the Kullback-Leibler divergence between our estimated density and the true density (Silverman, 1986). Furthermore, the likelihood score is naively computable in $O(n^2)$ time, which gives us a better starting point for deriving a fast approximation algorithm.

---

**Algorithm 1** Generic dual-tree recursion
　**Input:** nodes $r_i$, $r_j$; error tolerance $\epsilon$
　**if Can-approximate**$(r_i, r_j, \epsilon)$ **then**
　　**Approximate**$(r_i, r_j)$, **return**
　**end if**
　**if** leaf$(r_i)$ and leaf$(r_j)$ **then**
　　**DualtreeBase**$(r_i, r_j)$, **return**
　**else**
　　**Dualtree**$(r_i.lc, r_j.lc)$, **Dualtree**$(r_i.lc, r_j.rc)$
　　**Dualtree**$(r_i.rc, r_j.lc)$, **Dualtree**$(r_i.rc, r_j.rc)$
　**end if**

---

### 3.1 Dual-tree fast approximation

Dual-tree recursion is a spatial-partitioning approach for accelerating a variety of N-body computations such as kernel density estimates. We present a brief overview here and refer the reader to the original papers for greater detail (Gray & Moore, 2000; Moore et al., 2000; Gray & Moore, 2003).

For a double summation $\sum_i \sum_j g(x_i, x_j)$ over the data, the essential idea is that we can partition the set of pairs $(x_i, x_j)$ into subsets where the value $g(x_i, x_j)$ is approximately constant and can therefore be approximated once for the whole subset rather than explicitly computed for every pair.

Suppose the data $\{x_i\}$ are partitioned into subsets $r \in R$. We can write $\sum_i \sum_j g(x_i, x_j) = \sum_{r_i \in R} \sum_{r_j \in R} g(r_i, r_j)$, where $g(r_i, r_j) = \sum_{i \in r_i} \sum_{j \in r_j} g(x_i, x_j)$. If, for a given pair $(r_i, r_j)$, we can determine that $g(x_i, x_j)$ lies within sufficiently narrow bounds for all $x_i \in r_i$ and $x_j \in r_j$, then we can approximate by assuming all pairs in $(r_i, r_j)$ have some value in that range without calculating each term explicitly. The bounds on $g(x_i, x_j)$ can be used to bound the error caused by the approximation.

In a dual-tree recursion (see Algorithm 1) we produce partitions $R_i$ and $R_j$ over $\{x_i\}$ and $\{x_j\}$ by traversing two separate kd-trees.[1] $R_i$ and $R_j$ are initially the root nodes of the kd-trees; note that every kd-tree node provides a tight bounding box for the points it contains. Starting with the roots, each call to the algorithm examines a node pair $(r_i, r_j)$ to determine whether their contribution can be approximated. If so, we prune that branch of the recursion. If not, the two nodes are each replaced by their kd-tree children, resulting in four re-cursive node-node comparisons, unless both nodes are leaves, in which case further refinement is impossible and we are forced to do the brute-force computation for that pair.

---
[1] A kd-tree is a type of binary space partitioning tree used to speed up various kinds of computations; see (Moore et al., 2000) for details.

### 3.2 Fast cross-validated likelihood

We now derive a dual-tree-based approximation to the cross-validated likelihood $L$. First we note that, upon expansion of the $\hat{f}^{-i}$ terms, we can write

$$L = \frac{1}{n}\sum_i \log(\hat{f}^{-i}(y_i|x_i)\hat{f}^{-i}(x_i))$$

$$= \frac{1}{n}\sum_i \log(\frac{\sum_{j \neq i} K_{h_1}(y_i - y_j)K_{h_2}(||x_i - x_j||)}{\sum_{j \neq i} K_{h_2}(||x_i - x_j||)})$$

$$+ \log(\frac{1}{n-1}\sum_{j \neq i} K_{h_2}(||x_i - x_j||))$$

$$= \frac{1}{n}\sum_i \log(A_i) - \log(n-1) \ ,$$

where $A_i = \sum_{j \neq i} K_{h_1}(y_i - y_j)K_{h_2}(||x_i - x_j||)$. We will construct our approximation to $L$ by first constructing a list of approximations to the $A_i$ terms, then summing their logs to get $L$. In order to see how the various approximation errors will propagate through to the total error in $L$, we will make use of the following exact error propagation bounds, which we state as a lemma.

**Lemma 1** *The following rules hold for the absolute error $\triangle f$ induced in a function $f$ when operating on estimates of its arguments $x_i$ with absolute error no more than $\triangle x_i$:*

1. *If $f = \sum_i c_i x_i$, then $\triangle f \leq \sum_i c_i \triangle x_i$*
2. *If $f = \log(x)$, then $\triangle f \leq \log(1 + \frac{\triangle x}{|x|})$*

We want to guarantee that our approximation leaves us with $\triangle L \leq \epsilon$ for some user-specified error tolerance $\epsilon$. Using Rule 1, we have $\triangle L \leq \frac{1}{n}\sum_i \triangle \log(A_i)$, so our original error condition will hold if we can enforce $\frac{1}{n}\sum_i \triangle \log(A_i) \leq \epsilon$, which we can do by enforcing $\forall i, \triangle \log(A_i) \leq \epsilon$. Invoking Rule 2 and a bit of rearrangement, this is in turn equivalent to $\forall i, \frac{\triangle A_i}{A_i} < e^\epsilon - 1$.[2]

Now consider a partitioning $R$ of the index values $i$: we can write $A_i = \sum_{r \in R}\sum_{j \in r, j \neq i} v(i,j)$, where $v(i,j) = K_{h_1}(y_i - y_j)K_{h_2}(||x_i - x_j||)$. Then $\triangle A_i \leq \sum_r \triangle \sum_{j \in r, j \neq i} v(i,j)$, and our enforcement condition holds if $\frac{\sum_r \triangle \sum_{j \in r, j \neq i} v(i,j)}{\sum_r \sum_{j \in r, j \neq i} v(i,j)} \leq e^\epsilon - 1$, which is satisfied if $\forall r, \frac{\triangle \sum_{j \in r, j \neq i} v(i,j)}{\sum_{j \in r, j \neq i} v(i,j)} \leq e^\epsilon - 1$. We summarize this intermediate result as a lemma before describing two different ways of performing the approximation.

---

[2] We drop the absolute value signs on $A_i$ because, being a sum of kernel products, it is non-negative.

**Lemma 2** *If for each $A_i$ the contribution $\sum_{j \in r, j \neq i} v(i,j)$ from each subregion $r$ of a partitioning $R$ of the data can be approximated with relative error no greater then $e^\epsilon - 1$, then the total absolute error in $L$ computed by summing the logs of the approximated $A_i$ will be no greater than $\epsilon$.*

### 3.3 Approximation with deterministic bounds

Now consider two partitionings $R_i$ and $R_j$ induced by dual kd-tree traversals for the $i$ and $j$ indices. At any point in the dual-tree algorithm, $r_i$ and $r_j$ each contain some subset of the indices. The contribution from $r_j$ to the $A_i$ of each $i \in r_i$ is $\sum_{j \in r_j, j \neq i} v(i,j)$, and we can put bounds on the terms $v(i,j)$ for all $i \in r_i$ and $j \in r_j$ using the bounding boxes of $r_i$ and $r_j$. If the bounds on $v(i,j)$ are tight enough that we can construct an approximation satisfying Lemma 2, then we can trigger **Can-approximate** for these contributions while staying consistent with the global error bound, otherwise we must invoke the recursive comparison of the children of $r_i$ and $r_j$.

What we need, then, is an approximator $\hat{S}_r$ for $S_r = \sum_{j \in r, j \neq i} v(i,j)$, and an upper bound on its relative error. We give these in the following lemma, after which we can establish the correctness of the corresponding dual-tree pruning rule (i.e. the **Can-approximate** test in Algorithm 1).

**Lemma 3** *Let $v_r^{min}$ and $v_r^{max}$ be valid bounds for $v$ in the sum $S_r$. If we approximate $S_r$ by $\hat{S}_r = (n_r - 1)\hat{v}_r$, where $n_r$ is the number of points in $r$ and $\hat{v}_r = \frac{v_r^{max} + v_r^{min}}{2}$, the absolute error of the approximation is no greater than $(n_r - 1)\frac{v_r^{max} - v_r^{min}}{2} + v_r^{max}$.*

*Proof.* First note that setting $\hat{v}_r$ to the midpoint of $v_r^{max}$ and $v_r^{min}$ means that no value $v$ in $S_r$ can differ from $\hat{v}_r$ by more than $\frac{v_r^{max} - v_r^{min}}{2}$. With $\hat{S}_r = (n_r - 1)\hat{v}_r$, we have $\triangle S_r = |(n_r - 1)\hat{v}_r - \sum_{j \in r, j \neq i} v(i,j)| \leq \max\{(n_r - 1)\frac{v_r^{max} - v_r^{min}}{2} + v_r^{max}, (n_r - 1)\frac{v_r^{max} - v_r^{min}}{2}\}$. The first term in the max handles the case where $i \notin r$, in which case $\hat{S}_r$ matches all but one of the terms in $S_r$ with a $\hat{v}_r$. There is no more than $\frac{v_r^{max} - v_r^{min}}{2}$ error from each matched point, plus no more than $v_r^{max}$ from the single unmatched point. The second term in the max handles $i \in r$, in which case all terms in $S_r$ are matched by terms in $\hat{S}_r$. The first term in the max is at least as large as the second, and the lemma follows. □

**Theorem 1** *A dual-tree evaluation of the $A_i$ in L that approximates $S_r$ by $\hat{S}_r$ when $\frac{(n_r+1)v_r^{max}}{(n_r-1)v_r^{min}} \leq 2e^\epsilon - 1$ will guarantee that $\triangle L \leq \epsilon$.*

*Proof.* By Lemma 2, if all $S_r$ approximations satisfy $\frac{\triangle S_r}{S_r} \leq e^\epsilon - 1$, then $\triangle L \leq \epsilon$. Since $S_r \geq (n_r - 1)v_r^{min}$, we have $\frac{\triangle S_r}{S_r} \leq \frac{\triangle S_r}{(n_r-1)v_r^{min}}$, so we can enforce the condition of Lemma 2 by ensuring that $\frac{\triangle S_r}{(n_r-1)v_r^{min}} \leq e^\epsilon - 1$. By Lemma 3, approximating with $\hat{S}_r = (n_r - 1)\hat{v}_r$ gives $\triangle S_r \leq (n_r - 1)\frac{v_r^{max} - v_r^{min}}{2} + v_r^{max}$. Substituting this bound and rearranging shows that the Lemma 2 condition is implied by enforcing $\frac{(n_r+1)v_r^{max}}{(n_r-1)v_r^{min}} \leq 2e^\epsilon - 1$, which establishes the theorem.□

### 3.4 Approximation with probabilistic bounds

While dual-tree evaluation with this pruning rule guarantees the desired error bound, in practice it operates extremely conservatively because $v_r^{min}$ and $v_r^{max}$ are extreme values that may be very far from the majority of the terms being summed. To get a better idea of the composition of the values contained in a node pair, we employ a new sampling-based bootstrap approximation for the $\frac{\triangle S_r}{S_r} \leq e^\epsilon - 1$ test of Lemma 2. The procedure is defined in Algorithm 2.

We sample from the terms of $S_r$ for an estimate $\hat{v}$ of the mean value of $v$ over non-duplicate point pairs in $(r_i, r_j)$. We treat this as a sample statistic and obtain a bootstrap estimate of its variance $\hat{\sigma}_{\hat{v}}$. We then form a normal-based $1 - \alpha$ confidence interval $(\hat{v} \pm z_{\alpha/2}\hat{\sigma}_{\hat{v}})$[3] and estimate $\triangle S_r$ as $n_r z_{\alpha/2}\hat{\sigma}_{\hat{v}}$, $S_r$ as $n_r\hat{v}$, and therefore $\frac{\triangle S_r}{S_r}$ as $\frac{z_{\alpha/2}\hat{\sigma}_{\hat{v}}}{\hat{v}}$. This estimate is plugged into the condition of Lemma 2 to form a new approximate pruning rule. In practice, this allows for far more aggressive approximations and greatly speeds up the algorithm while still keeping error under control (overconservatism in the error propagation framework gives room for the increased error of this approach).

---

**Algorithm 2** Sample-based estimation of $\frac{\triangle S_r}{S_r}$

---
**Input:** nodes $r_i, r_j$; sample size $m$; bootstrap size $B$
samples ← ∅
**for** $k = 1$ **to** $m$ **do**
　Uniformly sample $s_k$ from $r_j$ until $s_k \notin r_i$
　samples ← samples $\cup s_k$
**end for**
$\hat{v}$ ← average(samples)
$\hat{\sigma}_{\hat{v}}$ ← bootstrapStdev(samples, B)
**return** $\frac{z_{\alpha/2}\hat{\sigma}_{\hat{v}}}{\hat{v}}$

---

[3]Because $\hat{v}$ is a sample mean, this is justified by the Central Limit Theorem; the sample size $m$ should be large enough to make the normal approximation valid.

### 3.5 Optimization

Due to the noisiness of $L$ over the bandwidth space, its gradient provides little utility above a random search. Thus, we select bandwidths by random sampling from a finite range large enough to cover the space of reasonable bandwidths, keeping the best ones according to the approximate evaluation of $L$.

## 4 Experiments

We present results from three classes of experiments. The first set is designed to test the efficiency and accuracy of our likelihood approximations on real datasets of varying size and dimension. The second set is composed of several synthetic datasets designed to allow direct evaluation of the estimated conditional density $\hat{f}(y|x)$ with respect to known generating distributions. The final experiment measures performance on a challenging redshift prediction task of active scientific interest from the Sloan Digital Sky Survey and compares our performance with that of a predictor from the astronomical community.

In all cases we use the asymptotically optimal Epanechnikov kernel. We scale the data by dividing each dimension by its standard deviation, which effectively allows $h_2$ to provide a different bandwidth $h_2\sigma_d$ for each dimension $d$ of $x$. Bandwidths are sampled from $[0, h_{max}]$ with $h_{max}$ no greater than 10, giving effective bandwidths from $[0, h_{max}\sigma_d]$ in each dimension. Prediction intervals of confidence $1 - \alpha$ are generated by taking a large sample from $\hat{f}(y|x)$, then quickly searching for the narrowest set of quantiles $q_{lower}, q_{lower+1-\alpha}$ that give the desired coverage.

### 4.1 Time Trials

In this set of experiments we ran bandwidth selection on various datasets to characterize the average training time and error of the naive, deterministic, and probabilistic likelihood scores. The first dataset, containing geyser duration and wait times (Bashtannyk & Hyndman, 2001), is two-dimensional, and we subsampled it at sizes 100, 200, and 299. The second dataset is subsampled from a set of three-dimensional Sloan Digital Sky Survey positional data, which we ran at sizes 500, 1000, 2000, and 10,000. Lastly, a nine-dimensional dataset containing a large set of neighborhood demographic census statistics was also run at sizes 500, 1000, 2000, and 10,000. Results are summarized in Figure 2 (all results appear on the last page, following the references).

Runtimes represent a full bandwidth selection, i.e. the total time required to evaluate a broad range of band-widths during the selection process, and are averaged across a 10-fold cross-validation. Errors are reported as the absolute deviation of the approximated values of $L$ from the exact values, and are averaged across 100 randomly selected bandwidth evaluations. Because of the intractability of exact likelihood evaluation, the larger datasets have no error measurements and their speedups are based on extrapolations of the base $O(N^2)$ runtime. Note that extremely small bandwidths cause $L$ to diverge, which allows us to short-circuit its computation, but these instances were thrown out of the averages in order to avoid downward skewing. In the probabilistic approximation of $L$, we used $m = 25$, $B = 10$, and $z_{\alpha/2} = 1.5$; performance was not extremely sensitive to the exact values, and these were chosen as representative of a good tradeoff between speedup and error.

The probabilistic approximation is one to two orders of magnitude faster than the deterministic approximation for datasets of up to 1000 points, while also remaining tractable on datasets in the low tens of thousands that the deterministic approximation is too slow to compute. Speedups for the probabilistic approximation reached around 6.1 thousand on the SDSS DR4 data and 3.8 million on the census data, and uniformly increased with sample size for each dataset. Values of $\epsilon$ for each method were calibrated to give an error of approximately 0.4 on the geyser dataset, then held at these values for all other datasets. The measured error shows two interesting trends: 1) it appears to slowly increase with the size of a given dataset, and 2) it appears to drop markedly, indicating increased overconservatism, as dimension increases. This dimension effect is acute in the deterministic case and less dramatic but still significant in the probabilistic case. This makes sense, as it becomes harder to affect min/max distances by splitting bounding boxes in higher dimensions, while a sample-based method more directly accesses the actual composition of the data. Having established that the probabilistic approximation is much faster, while still maintaining good error performance, we use the probabilistic approach exclusively in the remaining experiments.

### 4.2 Synthetic datasets

We now evaluate the performance of KCDE when using the probabilistic likelihood approximation for bandwidth selection. We evaluate with respect to three synthetic densities for which we know the underlying $f(y|x)$. As a measure of the density quality, we compute the mean $(\hat{f}(y_i|x_i) - f(y_i|x_i))^2$ over all points $(x_i, y_i)$ in the held-out sets of a 10-fold cross-validation. This is a consistent estimate of $\int (\hat{f}(y|x) - f(y|x))^2 f(y, x) dy dx$; we refer to it as ISE in the results, and its value goes to zero as $\hat{f}(y|x)$ approaches $f(y|x)$. Although KCDE is not designed specifically for regression, we also measure the mean squared error (MSE) of predicting $y_i$ by the expectation of $\hat{f}(y|x_i)$. Prediction intervals are generated, with quality measured by 1) the fraction of held-out points that fall within the intervals (Interval Coverage), and 2) the average half-widths of the intervals as a fraction of the true $y_i$ (Interval Width). As a baseline, we compare the metrics from our bandwidth selection method to those of a standard reference rule found in (Silverman, 1986). This reference rule is for standard kernel density estimation, and we use it on the $x$ and $y$ kernels separately as though they were unconditional estimators.

Table 1 summarizes these metrics on the three synthetic datasets. The first dataset, bimodal sine, is a typical regression setting where y is drawn from Gaussian noise around a deterministic function of x, in this case the sine function. We introduce bimodality by flipping the sign of the sine with 20% probability. The second dataset is drawn from a uniform distribution in five dimensions, with the width of the distribution in dimension $d$ equal to $2^d$. This moves us up in dimension while providing a very different setting from the bimodal sine because of the need for wide kernels. Lastly, the decay series is a time series in which the value at a given timestep is sampled from a unit Gaussian centered on one of the seven previous timesteps with exponentially decaying probability. This can be thought of as a random walk with a bit of momentum, and is designed to demonstrate the applicability to time series and to give a higher-dimensional test, with the vectors $x$ composed of the seven previous observations at each timestep $y$.

As the table data indicate, our data-driven method clearly dominates the reference rule in terms of ISE and MSE. The prediction intervals were generated at 95%, and the likelihood-based density manages coverage more consistently near that mark. The likelihood metrics are much less variable than those of the reference rule: in the few instances where the reference rule obtains a better metric value, the likelihood is close behind, but when likelihood achieves the better mark the reference rule is often drastically worse.

These synthetic results validate likelihood-based bandwidth selection relative to a sane baseline, justifying the use of this approach on real datasets.

### 4.3 SDSS redshift

One task in the Sloan Digital Sky Survey is to determine the redshifts of observed astronomical objects, from which their distances can be computed in order to map out cosmic geometry. Redshifts can be determined accurately from expensive spectroscopic measurements, or they can be estimated from cheaper optical measurements that roughly approximate the complete spectra. This sounds like a regression task, but often what appear to be single objects are in fact combinations of sources at different distances. Thus, a given set of optical measurements can yield a multimodal distribution over redshifts, which is well modeled by conditional density estimation. Multi-modality and wide prediction intervals could be used as triggers for engaging the finer measurements.

We present results from applying likelihood-based KCDE to samples of 10,000 and 20,000 objects from the SDSS. Each data point consists of four spectral features plus the true redshift/distance measurement. Metrics are the same as for the synthetic data, aside from the ISE which we can no longer compute with respect to a known underlying distribution. We again compare to the reference rule baseline, with results summarized in Table 2. Prediction interval coverage is much better with the likelihood criterion. MSE is about 8% better with the likelihood criterion at $n = 10K$, and 2% worse at $n = 20K$. These MSE values are only about 30% higher than the .743 posted by the astronomers' custom regression function. This is good, given that our estimator is not optimizing for regression accuracy and makes no parametric or domain-knowledge assumptions about the problem.

In sum, these empirical results collectively validate the speedup and error control of our fast likelihood approximations, while demonstrating the effectiveness of likelihood-based bandwidth selection, which appears to work well across different kinds of data. The probabilistic likelihood approximation is orders of magnitude faster than the deterministic approximation while still keeping error under control. Likelihood-based bandwidth selection outperformed a baseline reference rule, and the conditional density estimates demonstrated their versatility by performing well under a variety of metrics.

## 5 Summary and future work

We have described kernel conditional density estimation, a data modeling approach with various uses from visualization to prediction. We defined a likelihood cross-validation bandwidth score and showed how to approximate it efficiently using both deterministic and probabilistic error control. Speedups ranged from 50 to 3.8M with the probabilistic approximation, and from 1.25 to 667 with the deterministic, always increasing, for a given dataset, with the number of points. These fast likelihood computations have the potential to be used in other areas such as nonparametric Bayesian inference. We demonstrated good performance of likelihood-based KCDE on a variety of datasets, including some that were larger by an order of magnitude than any previous applications of this type of estimator.

By no means have we exhausted the potential for speeding up bandwidth selection. Nor have we come near to determining all the machine learning applications of the kernel conditional estimator, but we believe it is a fundamental technique that will have widespread applicability, as described in the introduction. We intend to explore such applications, along with increased efficiency gains, much larger problems, and speedup of the more robust ISE criterion.

## References

Bashtannyk, D. M., & Hyndman, R. J. (2001). Bandwidth selection for kernel conditional density estimation. *Computational Statistics & Data Analysis*, *36*, 279–298.

Fan, J., Yao, Q., & Tong, H. (1996). Estimation of conditional densities and sensitivity measures in nonlinear dynamical systems. *Biometrika*, *83*, 189–206.

Fan, J., & Yim, T. H. (2004). A crossvalidation method for estimating conditional densities. *Biometrika*, *91*, 819–834.

Gooijer, J. G. D., & Zerom, D. (2003). On conditional density estimation. *Statistica Neerlandica*, *57*, 159–176.

Gray, A. G., & Moore, A. W. (2000). N-body problems in statistical learning. *Advances in Neural Information Processing Systems (NIPS) 13*.

Gray, A. G., & Moore, A. W. (2003). Rapid evaluation of multiple density models. *Workshop on Artificial Intelligence and Statistics*.

Hansen, B. E. (2004). Nonparametric conditional density estimation. Unpublished manuscript.

Hyndman, R. J., Bashtannyk, D. M., & Grunwald, G. K. (1996). Estimating and visualizing conditional densities. *Journal of Computation and Graphical Statistics*, *5*, 315–336.

Moore, A., Connolly, A., Genovese, C., Gray, A., Grone, L., Kanidoris II, N., Nichol, R., Schneider, J., Szalay, A., Szapudi, I., & Wasserman, L. (2000). Fast algorithms and efficient statistics: N-point correlation functions.

Rosenblatt, M. (1969). Conditional probability density and regression estimators. In P. R. Krishnaiah (Ed.), *Multivariate analysis ii*, 25–31. Academic Press, New York.

Schapire, R. E., Stone, P., McAllester, D., Littman, M. L., & Csirik, J. A. (2002). Modeling auction price uncertainty using boosting-based conditional density estimation. *ICML*.

Silverman, B. W. (1986). *Density estimation for statistics and data analysis*, vol. 26 of *Monographs on Statistics and Probability*. Chapman & Hall.

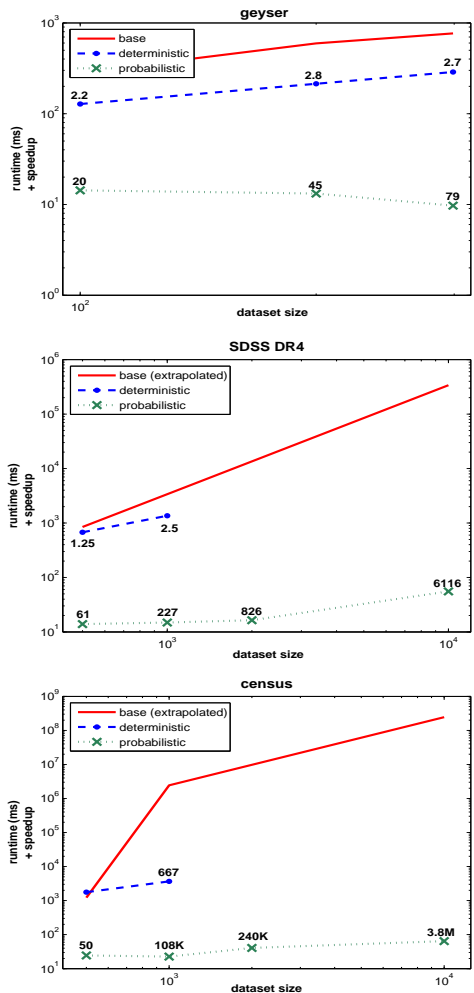

| **geyser** | error | | |
|---|---|---|---|
| | n=100 | n=200 | n=299 |
| deterministic | .036 | .046 | .038 |
| probabilistic | .036 | .038 | .042 |

| **SDSS DR4** | error | | | |
|---|---|---|---|---|
| | n=500 | n=1000 | n=2000 | n=10K |
| deterministic | .0089 | .0081 | - | - |
| probabilistic | .053 | .066 | .065 | - |

| **census** | error | | | |
|---|---|---|---|---|
| | n=500 | n=1000 | n=2000 | n=10K |
| deterministic | 0 | 0 | - | - |
| probabilistic | .00046 | .00078 | .00087 | - |

Figure 2: Average runtime and absolute error vs. dataset size for each computation method on the 2-dimensional geyser data, 3-dimensional SDSS DR4 data, and 9-dimensional census demographic data. Speedups are annotated above each point on the runtime graphs, which are log-log scaled. Speedups for $n > 500$ are relative to a quadratic extrapolation of the naive runtime.

Table 1: Performance of fast likelihood and reference rule selection criteria on synthetic datasets. The better of each pair of metrics is in bold face, except when interval widths are not comparable due to disparity in the coverage rates.

| | (fast likelihood/reference rule) | | | |
|---|---|---|---|---|
| Dataset/size | ISE | MSE Expectation | Interval Coverage | Interval Width |
| bimodal sine/10K | **.352**/.720 | **3.02**/3.27 | .972/**.958** | 4.84/**3.69** |
| uniform/15K | **.208**/.755 | **51.0**/56.2 | **.996**/.421 | 7.12/2.25 |
| decay series/2K | **1.39**/1.89 | **1.63**/62.6 | .992/**.991** | **5.12**/12.1 |

Table 2: Performance of fast likelihood and reference rule selection criteria on the SDSS redshift dataset. The better of each pair of metrics is in bold, except when interval widths are not comparable due to disparity in the coverage rates.

| | (fast likelihood/reference rule) | | |
|---|---|---|---|
| Dataset/size | MSE Expectation | Interval Coverage | Interval Width |
| SDSS redshift/10K | **.975**/1.05 | **.969**/.800 | .974/.400 |
| SDSS redshift/20K | .980/**.959** | **.980**/.816 | 1.09/.408 |